\documentclass{PoS}

\usepackage{txfonts}
\usepackage{bm}
\newcommand{\BB}{{\bf B}}
\newcommand{\EE}{{\bf E}}
\newcommand{\UU}{{\bf U}}
\newcommand{\JJ}{{\bf J}}
\newcommand{\bb}{{\bf b}}
\newcommand{\jj}{{\bf j}}
\newcommand{\uu}{{\bf u}}
\newcommand{\aaa}{{\bf a}}
\newcommand{\meanUU}{\overline{\mbox{\bf U}}}
\newcommand{\meanBB}{\overline{\mbox{\bf B}}}
\newcommand{\meanJJ}{\overline{\mbox{\bf J}}}
\newcommand{\meanAA}{\overline{\mbox{\bf A}}}
\newcommand{\meanFF}{\overline{\mbox{\bf F}}}
\newcommand{\meanEMF}{\overline{\mbox{\boldmath ${\mathcal E}$}} {}}
\newcommand{\bra}[1]{\langle #1\rangle}
\newcommand{\nab}{{\bf \nabla}}
\newcommand{\oo}{{\bf \omega}}

\newcommand{\Rm}{R_\mathrm{m}}
\newcommand{\Rmcr}{R_\mathrm{m,cr}}
\newcommand{\Rey}{\mathrm{Re}}
\newcommand{\Pm}{P_\mathrm{m}}
\newcommand{\EQ}{\begin{equation}}
\newcommand{\EN}{\end{equation}}
\newcommand{\EQA}{\begin{eqnarray}}
\newcommand{\ENA}{\end{eqnarray}}

\title{Magnetizing the universe}

\ShortTitle{Magnetizing the universe}

\author{\speaker{Kandaswamy Subramanian}%
	 \\
        Inter University Centre for Astronomy and Astrophysics, Post Bag 4, Ganeshkhind, Pune-411007, India. \\
        E-mail: \email{kandu@iucaa.ernet.in}}

\abstract{
The origin of cosmic magnetism is an issue of fundamental importance
in astrophysics. We review here some of the ideas of how
large scale magnetic fields in the universe, 
particularly in galaxies and galaxy clusters could arise.  
The popular paradigm involves the generation of a seed magnetic field
followed by turbulent dynamo amplification of the seed field. We first
outline various seed field generation mechanisms including Biermann
batteries. These in general give
a field much smaller than the observed field and so they require further
amplification by dynamo action. The basic idea behind fluctuation dynamos,
as applied to cluster magnetism and the mean-field helical dynamo as applied
to disk galaxies, are outlined.  Major difficulties with the dynamo paradigm
are considered. It is particularly important 
to understand the nonlinear saturation of dynamos, and whether the fields
produced are coherent enough on large-scales to explain the observed fields
in galaxies and clusters. At the same time the alternative possibility of
a primordial field lacks firm theoretical support but can have very interesting
observational consequences.
}

\FullConference{From Planets to Dark Energy: the Modern Radio Universe\\
		 October 1-5 2007\\
		 The University of Manchester, UK}

\begin{document}

\section{Introduction}

Magnetic fields are crucial for understanding a number of physical processes
in the universe. They give "fluid" like properties to hot and dilute plasma and
affect transport properties like thermal
conduction, viscosity and resistivity of plasmas. 
Present day star formation is controlled by magnetic fields
with the field possibly determining the typical
masses of stars and the initial mass function. They are important for
accretion and ejection flows from protostars to active galactic nuclei (AGN).
The galactic field is important for cosmic ray confinement and transport.
Their most important role for radio astronomers is that
they make it possible to "see" objects due to synchrotron and
inverse Compton emission of relativistic electrons gyrating in
the magnetic field. Indeed it is this radio emission that is
crucial to measure the strengths and coherence scales of magnetic
fields on the largest scales in the universe, that
associated with galaxies and galaxy clusters \cite{Beck00}. 
Such studies show average total field strengths in nearby spiral galaxies
$B \sim 10 \mu$G with a mean field about half this value, correlated
on scales of several kpc to tens of kpc.
Galaxy clusters also seem to host magnetic fields of several $\mu$G, and
correlated on several kpc scales.
How do such ordered large-scale fields arise in galaxies and clusters? 
Indeed the origin cosmic magnetism in general is one of the fascinating questions
of modern astrophysics. It forms one of the key goals of 
the Square Kilometre Array (SKA). 
We review here some of the ideas that have been put forth on how
the universe got magnetized. 

We start with a brief outline of the basics of magnetohydrodynamics,
consider then both seed field generation mechanisms and
dynamo amplification processes.
We consider the possibility of magnetizing the universe
through outflows from starbursting galaxies and AGN.
We finally discuss the possible role of primordial magnetic fields
created perhaps in an early universe phase transition.
We focus on magnetic fields on scales larger than galaxies
here, as such fields will also be the focus of SKA type telescopes,
and do not discuss stellar or planetary magnetism. 
An extensive review 
of several aspects of astrophysical magnetic fields and nonlinear dynamos
can be found in \cite{BS05a}.

\section{Basic MHD}

The evolution of the magnetic field $\BB$ is governed by the 
induction equation
\EQ
\frac{\partial \BB}{\partial t}
=\nabla\times\left({\bf U}\times{\bf B} -\eta\nabla\times{\bf B} \right)
\label{induc}
\EN
got from combining Maxwell equations (neglecting the displacement current) and
a simple form of Ohms law. Here $\UU$ is the fluid velocity and $\eta$ the
microscopic resistivity.
The field back reacts on ${\bf U}$ via the Lorentz force ${\bf J} \times {\bf B}$
\EQ
\rho\left[\frac{\partial \UU}{\partial t} + (\UU\cdot\nab)\UU \right]
=-{\bf \nabla} p+\frac{{\bf J} \times{\bf B}}{c}
+ {\bf f}+ {\bf F}_{\rm visc},
\EN
where $\rho$, $p$ are the fluid density and pressure, 
${\bf f}$ represents all the body forces and ${\bf F}_{\rm visc}$
the viscous force.
A few simple consequences of the above equations are as follows:
(i) If ${\bf U}=0$, then the field simply decays due to finite resistivity.
(ii) On the other hand for $\eta \to 0$, the magnetic flux
$\Phi = \int_S {\bf B}\cdot d{\bf S}$ through any surface moving
with the fluid is 'frozen' in the sense that $d\Phi/dt \to 0$.
(iii) The magnetic Reynolds number $\Rm = (UB)/(\eta B/L) = UL/\eta $
measures the relative importance of the induction and resistive effects. 
(Here $U$ and $L$ are respectively the typical velocity and length scale 
in the system.) 
For most astrophysical systems $\Rm \gg 1$.
In galaxies we expect naively $\Rm \sim 10^{18}$ and in clusters 
$\Rm \sim 10^{29}$ based on the Spitzer resistivity \cite{BS05a}.
Plasma instabilities in the presence of a weak magnetic field
could produce fluctuating magnetic fields which scatter charge
particles and affect various transport processes \cite{scheko05}. 
But to affect resistivity by scattering electrons, these fluctuations need
to have significant power down to the electron gyro radius.
(iv) Note that ${\bf B} =0$ is perfectly valid solution to
the induction equation! So magnetic fields can only 
arise from a zero field initial condition, if the usual form of Ohm's law is violated
by a "battery term" (see below), to generate a seed magnetic field $B_{seed}$.
(v) The resulting seed magnetic field is generally very
much smaller than observed fields in galaxies and clusters and
so one needs the motions (${\bf U}$) to act as a dynamo
and amplify the field further.
It turns out to be then crucial to understand how dynamos work and saturate.

At this stage it is important to clarify the following: it is often mistakenly 
assumed that if one has a pre-existing magnetic field in 
a highly conducting medium, and the resistive decay time is longer than say
a Hubble time, then one does not need any mechanism for maintaining such
a field. This is not true in general; because given a tangled field the
Lorentz forces would drive motions in the fluid. These would either
dissipate due to viscous forces if such forces were important
(at small values of the fluid Reynolds number $\Rey=UL/\nu$, where $\nu$
is the fluid viscosity) or if viscosity is small, drive 
decaying MHD turbulence, with a cascade of energy to smaller and smaller scales and
eventual dissipation on the dynamical timescales associated with the motions. 
This timescale can be much smaller than the age of the system. 
For example if clusters host a few $\mu$G magnetic fields say tangled on 
say $l\sim 10$ kpc scales, such fields could typically decay on the
Alfv\'en crossing timescale $l/V_A \sim 10^8$yr, where we have
taken a typical Alfv\'en velocity $V_A \sim 100$ km s$^{-1}$. 
Although the energy density in MHD turbulence decays with time
as a power law, this time scale is still much shorter than the
typical age of a cluster, which is thought to be several billion years.
Similarly, if the fluid is already turbulent, the associated larger turbulent
resistivity will can lead to the decay of large-scale fields (except
perhaps in the presence of strong shear, where there may be dynamo action; see below).
Therefore, one has to provide explicit explanation of the
origin and persistence of cosmic magnetic
fields; reference to the low Ohmic resistivity of the
plasma is not sufficient if the gas is turbulent or the
magnetic field is tangled.
Hence even if the medium were highly conducting, in most cases, a dynamo
is needed to maintain the observed magnetic field.

\section{The first seed fields}
 
We concentrate on astrophysical batteries here and discuss primordial
magnetic fields arising from the early universe in Section 7. 
The basic problem any battery has to address is how to
produce finite currents from zero currents?
Most astrophysical mechanisms use the fact that positively and
negatively charged particles in a charge-neutral universe, do not
have identical properties. For example, if one considered a gas
of ionized hydrogen, then the electrons have a much smaller
mass compared to protons. Thus that for a given pressure gradient
of the gas the electrons tend to be accelerated much more than the ions.
This leads in general to an electric field, which couples back
positive and negative charges. If such a thermally generated electric
field has a curl, then from Faraday's law of induction a magnetic
field can grow. 
The resulting battery effect, known as the Biermann battery,
was first proposed as a mechanism for the
thermal generation of stellar magnetic fields \cite{Bier50}.

The thermally generated electric field is given by 
$\EE_{bier} = -{\bf \nabla} p_e /e n_e $ got by
balancing the forces on the electrons, due to pressure gradient
and the electric field and assuming the protons are much more
massive than the electrons. 
The curl of this term leads to an extra source term
in the 
induction equation, which if we adopt 
$p_e = n_e k_{\rm B}T$, where $k_{\rm B}$
is the Boltzmann constant, 
gets modified to,
\EQ
{\partial \BB \over \partial t} =
\nabla\times\left({\bf U}\times{\bf B} -\eta\nabla\times{\bf B} \right)
-{c k_{\rm B} \over e} {\nab n_e\over n_e}\times\nab T.
\label{modb}
\EN
Therefore over and above the usual
flux freezing and diffusion terms we have a {\it source term}
which is nonzero if and only if the density and
temperature gradients, $\nab n_e$ and $\nab T$, are not
parallel to each other.
In the cosmological context, such non-parallel density and
temperature gradients can arise in a number of ways.
For example, in cosmic ionization fronts 
produced when the first ultraviolet photon sources,
like starbursting galaxies and quasars, 
turn on to ionize the intergalactic medium (IGM),
the temperature gradient is 
normal to the front.  However, a component to the
density gradient can arise in a different direction, if the ionization
front is sweeping across arbitrarily laid down density fluctuations,
which will later collapse to form galaxies and clusters.
Such density fluctuations, will 
in general have no correlation to the source of the
ionizing photons, resulting in a thermally generated
electric field which has a curl, and magnetic fields
correlated on galactic scales can grow. After compression during
galaxy formation, they turn out to have a strength
$B \sim 3 \times 10^{-20}$G \cite{Sub94}. 
This scenario has in fact been confirmed in detailed numerical
simulations of IGM reionization \cite{gnedin00}.
The Biermann battery has also been shown to generate both
vorticity and magnetic fields in oblique cosmological shocks which arise
during cosmological structure formation
\cite{kulsrud97}.

The asymmetry in the mass of the positive and negative charges
can also lead to battery effects during the interaction 
of radiation with ionized plasma. Note that the Thomson cross
section for the scattering of photons with charged particles depend inversely on the
mass of the particle. So the electron component of an ionized plasma 
is more strongly coupled with radiation than the proton
component. Suppose one has a rotating fluid element in the presence of a 
radiation bath. The interaction with photons
will brake the velocity of the electron component faster than the proton component and
set up a relative drift and hence lead to magnetic field 
generation \cite{harrison}. 
In the modern context, second order effects during
recombination also leads to both vorticity and magnetic field
generation due to the $\gamma-e/p$ scattering asymmetry. The resulting
magnetic fields are again very small $B \sim 10^{-30}$G on Mpc scales
upto $B \sim 10^{-21}$G at parsec scales \cite{gopal-sethi}.

As can be gleaned from the few examples considered above all battery mechanisms
give only very small fields on the cosmological scales much smaller
than the observed fields.  One therefore needs some form of dynamo action to 
amplify these fields further. Of course, seed fields
for dynamos in larger scale objects like galaxies and clusters can
arise not necessarily due to battery effects, but also due to more
rapid dynamo generation in objects with a much shorter dynamical timescale,
like stars and AGN, and subsequent ejection of these fields in to the interstellar
and intra cluster media (cf. \cite{Rees87} for reviews). 
In this case much larger seed fields say for the galactic dynamo $B \sim 10^{-9}$G
are possible \cite{RSS88}.
The down side is the as yet unresolved question as to how magnetized
gas say ejected in a supernovae or AGN is mixed
with unmagnetized gas in the protogalaxy, and how this mixing
affects the coherence scale of the field? 
And even in this case one still needs dynamos to work efficiently
in stars and AGN to generate the seed field, and in galaxies
and clusters to further amplify and maintain it against decay.  

\section{Fluctuation dynamos in galaxies and clusters}

Magnetic fields in a conducting medium can be amplified 
by the inductive effects associated with the motions of 
the medium. In this process, generally
referred to as a dynamo, the kinetic energy associated
with the motions is tapped to amplify magnetic energy. 
The plasmas in galaxies and clusters are most often turbulent. The dynamo in
this context is referred to as a `turbulent dynamo', and its analysis must
rely on statistical methods or direct numerical simulations. 
Turbulent dynamos are conveniently divided into 
fluctuation (or small-scale) and mean-field (or large-scale) dynamos. 
The fluctuation dynamo produces magnetic fields that are correlated
only on scales of the order of or smaller than the energy-carrying scale of
the random motions. We will discuss this here and turn to mean-field 
dynamos in the next section.

The importance of fluctuation dynamos in cosmic objects obtains
because they are generic in any random flow
where $\Rm$ exceeds a modest critical value $\Rmcr \sim 100$.
Fluid particles in such a flow randomly walk away from each other. A
magnetic field line frozen into such a fluid will then be extended by the
random stretching (if $\Rm$ is large enough). 
Consider a small segment of a thin flux tube of length $l$ and cross-section $A$,
and magnetic field strength $B$, in a highly
conducting fluid. Then, as the fluid moves about, conservation
of flux implies $BA$ is constant, and conservation of mass
implies $\rho A l$ is constant, where $\rho$ is the local
density. So $B \propto \rho l$. For a nearly incompressible
fluid, or a flow with small changes in $\rho$, one
will obtain $B \propto l$. Any random shearing motion which increases
$l$ will also amplify $B$; an increase in $l$ leading to
a decrease in $A$ (because of incompressibility) and hence
an increase in $B$ (due to flux freezing). 

Of course since the scale of
individual field structures decreases, (that is since $A \sim 1/B$), 
as the field strength increases, 
the rate of Ohmic dissipation increases until it compensates the effect of
random stretching. For a single scale random flow, this happens 
when $v_0/l_0 \sim \eta/l_B^2$, where $v_0$ is the typical velocity 
variation on scale $l_0$ and $l_B$ is the scale
of the magnetic field. This gives $l_B = l_\eta \sim l_0/\Rm^{1/2}$, 
the resistive scale $l_\eta$ for the flow, where $\Rm = v_0l_0/\eta$.
What happens after this can only be addressed by
a quantitative calculation. For a random flow which is delta-correlated in time,
it was shown by Kazantsev \cite{Kaz68}, that magnetic field
can grow provided $\Rm > \Rmcr\sim 30-100$, depending on the form of
the velocity correlation function.
In the kinematic regime the field grows exponentially
roughly on the eddy turnover time $l_0/v_0$ and
is also predicted to be intermittent,
i.e., concentrated into structures whose size, in at least one dimension,
is as small as the resistive scale $ l_\eta$ (e.g., \cite{RSS88,Zel90}).
In Kolmogorov turbulence, where the velocity variations on a
 scale $l$ is $v_l\propto l^{1/3}$, the $e$-folding time is shorter at
smaller scales, $l/v_l\propto l^{2/3}$, and so smaller eddies amplify the field
faster. If $\Pm = \Rm/\Rey \gg 1$, as relevant for galactic
and cluster plasma, even the viscous scale
eddies can exponentially amplify the field.

The fluctuation dynamo has since been convincingly confirmed beyond 
the limits of Kazantsev's theory by numerical simulations of forced and
flows \cite{Haugen,Schek04},
especially when $\Rm \ge \Rey$. Such simulations 
are also able to follow the fluctuation dynamo
into the non-linear regime where the Lorentz forces becomes strong enough to
affect the flow as to saturate the growth of magnetic field.

In the context of galaxy clusters turbulence would mainly be driven by
the continuous and ongoing merging of subclusters to form larger and
larger mass objects. One typically expects the largest turbulent scales
of $l_0 \sim 100$kpc and turbulent velocity $v_0 \sim 300$ km s$^{-1}$,
leading to a growth time $\tau_0 \sim l_0/v_0 \sim 3 \times 10^8$ yr; 
thus for a cluster lifetime of a few Gyr, one could then have significant
amplification by the fluctuation dynamo (cf. \cite{SSH06}).
And in the case of galactic interstellar turbulence driven
by supernovae, if we adopt values corresponding to the local ISM of 
$l_0 \sim 100$pc, $v_0 \sim 10$km s$^{-1}$ one gets $\tau_0 \sim 10^7$ yr.
Again the fluctuation dynamo would rapidly grow the magnetic field
even for very young high redshift protogalaxies.

The major uncertainty, in case we want to
use the fluctuation dynamo to explain observed Faraday rotation
measures in a galaxy clusters or that inferred in some high redshift
protogalaxies, is how intermittent the
field remains when the dynamo saturates. 
A simple model exploring an ambipolar drift type nonlinearity 
\cite{S99}, suggests that the
smallest scale of the magnetic structures will be renormalized in the
saturated state to become $l_B\simeq l_0\Rmcr^{-1/2}$
instead of the resistive scale $l_\eta$.
This essentially happens via a renormalization
of the effective magnetic diffusivity in these models \cite{S99},
such that the dynamo is 
saturated via a reduction of the effective magnetic Reynolds 
number down to its critical value for the dynamo action.
In this case one could indeed obtain significant Faraday rotation measure
through such a fluctuation dynamo generated field.
On the other hand it has been argued that 
the fluctuation dynamo generated field remains highly intermittent
even at saturation, with field reversals typically occurring on the resistive 
scale $l_\eta$ \cite{Schek04}.
Since the cluster $\Rm \sim 10^{29}$ if one uses naively
the collisional Spitzer resistivity, one hardly expects to
see any Faraday rotation from such a field.
Plasma effects would then important to renormalize the
effective $\Rm$, much below this ridiculously large value \cite{scheko05}.
One needs a better theoretical understanding 
of the non-linear saturation of fluctuation dynamos and the extent to
which plasma effects alter the resistivity of cluster plasma, to make
further progress.

Meanwhile direct simulations of the fluctuation dynamo at modest
values of $\Rm \sim 400-1000$, to examine
the resulting Faraday rotation, have been carried out \cite{SSH06}, 
following the methods of Haugen, Brandenburg and Dobler \cite{Haugen}. 
The simulations use a Cartesian box on a
cubic grid with $256^3$ mesh points.
The computational box contained just a few turbulent cells 
on the forcing scales.
The Faraday rotation measure $K\int n_e \BB\cdot d{\bf l}$ for
$256^2$ lines of sight through the computational domain was calculated,
and various properties of its distribution were determined including
the normalized standard deviation $\sigma_{RM}$,
normalized by $K \bar{n}_e B_{\rm rms} l_0$
(Here $n_e$ and $\bar{n}_e$ are the actual and average 
electron densities respectively, $B_{\rm rms}$ the RMS
value of the magnetic field in the box, $l_0$ the forcing scale
of the random flow, and $K$ 
is a constant of proportionality which converts the line
of sight integral into physical units of RM).
For a random field one expects a distribution of Faraday rotation
with mean zero and a standard deviation $\sigma_{RM}$, 
which depends on the degree of the field intermittency. 
If the field saturated with many resistive scale reversals
one expects $\sigma_{RM} \ll 1$, whereas if $l_B\simeq l_0\Rmcr^{-1/2}$ 
decides the coherence scale of the field, 
one expects $\sigma_{RM}$ of order unity. 
The simulations found the normalized $\sigma_{RM} \sim 0.5$ for
$\Pm=1, \Rey=\Rm\approx420$ and $\sigma_{RM} \sim 0.3$ for a $\Pm=30,
\Rey\approx44, \Rm\approx1300$ case.
Therefore the fluctuation dynamo does indeed seem to
be able to produce magnetic fields which will give significant
Faraday rotation measure, although simulations
at much higher values of $\Rm,\Rey$ are required.
Further, as emphasized by \cite{SSH06}, during epochs 
when cluster turbulence decays, one can still get a significant
$\sigma_{RM}$, as it decreases in time as a shallower power law than
the field itself, because of an increase in the field coherence scale.
Similar ideas involving cluster turbulence and fluctuation dynamos
have also been invoked by \cite{vogt_ensslin06} to explain
magnetic fields in cool cores of galaxy clusters.
The way the fluctuation dynamo saturates will be important
not only to decide if it can result in observable Faraday
rotation, but also to issues of Cosmic ray confinement in the
early protogalaxy.

\section{Mean field dynamos and galactic magnetism}

A remarkable change in the turbulent dynamo action
occurs if the turbulence is helical.
This can be clearly seen for example in the simulations
by Brandenburg \cite{B01}, where a large scale field, on the
scale of the box develops when a helical forcing is employed,
even though the forcing itself is on a scale about 1/5th the size of the box.
The large scale field however in these closed
box simulations develops only on the long resistive timescales.
It is important to understand how such a field develops and
how one can generate a large scale field on a faster timescale.
The possible importance of helical turbulence for large-scale
field generation was proposed by Parker \cite{Par55},
and is in fact discussed in text books \cite{Mof78}.
We summarize briefly below the theory of the mean-field
dynamo (MFD) as applied to magnetic field generation in
disk galaxies, turn to several potential problems
that have been recently highlighted and their possible resolution.

Suppose the velocity field is split into 
the sum of a mean, large-scale velocity $\meanUU$
and a turbulent, stochastic velocity $\uu$.
The induction equation becomes a stochastic
partial differential equation.
Let us split the magnetic field also as $\BB = \meanBB + \bb$, 
into a mean field $\meanBB = \bra{\BB}$ and a
fluctuating component ${\bf b}$. Here the average $<>$, is defined
either as a spatial average over scales larger than the turbulent
eddy scales (but smaller than the system size) or as an ensemble average.
Taking the average of the induction equation \ref{induc}, one gets
the mean-field dynamo equation for $\meanBB$,
\begin{equation}
\frac{\partial \meanBB}{\partial t} =
{\bf \nabla } \times \left( \meanUU \times \meanBB +
\meanEMF - \eta \nab \times \meanBB \right) .
\label{MFDeqn}
\end{equation}
This averaged equation now has a new term,
the mean electromotive force (emf)
$\meanEMF={\overline {\uu\times\bb}}$,
which crucially depends on the statistical properties of 
the small-scale velocity and magnetic fields.
A central closure problem in MFD theories is
to compute the mean emf $\meanEMF$
and express it in terms of the mean field itself.
In the two-scale approach one assumes that the mean field is spatially
smooth over scales bigger than the turbulence coherence scale $l$, 
and expresses the mean emf $\meanEMF$ in terms of the
mean magnetic field and its first derivative \cite{Mof78}.
For isotropic, homogeneous, helical 'turbulence'
in the approximation that the correlation time
$\tau$ is short (ideally $u\tau/l \ll 1$, where $u$ is the 
typical turbulent velocity) one employs what is known as the
First order smoothing approximation (FOSA) to write 
\begin{equation}
\meanEMF = \overline{\uu \times \bb}
= \alpha_K \meanBB - \eta_t \nab\times \meanBB.
\end{equation}
Here $\alpha_K = -(\tau/3)\overline{\oo\cdot\uu}$
is the dynamo $\alpha$-effect, proportional to the
kinetic helicity and $\eta_t= (\tau/3)\overline{\uu^2}/3$ 
is the turbulent magnetic diffusivity proportional to the kinetic energy
of the turbulence \cite{Mof78}.

In the context of disk galaxies, the mean velocity $\meanUU$ is that
of differential rotation. This leads to the $\Omega$-effect, that of
shearing radial fields to produce toroidal fields,
while the $\alpha$-effect is crucial for regeneration
of poloidal from toroidal fields.
A physical picture of the $\alpha$-effect is as follows:
The interstellar medium is assumed
to become turbulent, due to for example the effect of supernovae randomly
going off in different regions. In a rotating, stratified
(in density and pressure) medium
like a disk galaxy, such turbulence becomes helical.
Helical turbulent motions of the gas
perpendicular to the disk draws out the toroidal field
into a loop which looks like a {\it twisted} $\Omega$.
Such a twisted loop is connected to a current
which has a component parallel to the original toroidal
field. If the motions
have a non-zero net helicity, this
parallel component of the current adds up coherently.
A toroidal current then results from the toroidal field.
Hence, poloidal fields can be generated from toroidal ones.
(Of course microscopic diffusion is essential
to make permanent changes in the field).
This closes the toroidal-poloidal cycle and leads
to exponential growth of the mean field.
The turbulent diffusion turns out to be also essential for allowing changes
in the mean field flux. 
The kinematic MFD equation \ref{MFDeqn},
gives a mathematical foundation to the above
picture. One finds exponentially growing solutions, of
the MFD equations provided
a dimensionless dynamo number has magnitude
$D = \vert \alpha_0 G h^3 \eta_t^{-2} \vert >
D_{crit} \sim 6$ \cite{RSS88,Shuk}, a condition which can be satisfied
in disk galaxies.
(Here $h$ is the disk scale height and $G$ the galactic shear,
$\alpha_0$ typical value of $\alpha$, 
and we have defined $D$ to be positive).
The mean field grows
typically on time-scales a few times the rotation time scales,
of order $3-10 \times 10^8$ yr. 

This picture of the galactic dynamo faces several potential problems.
Firstly, while the mean field dynamo operates to generate
the large-scale field, the fluctuation dynamo is producing small-scale
fields at a much faster rate. Also the correlation time of
the turbulence measured by $u\tau/l$ is not likely to be
small in turbulent flows. So the validity of FOSA is questionable.
Indeed, based on specific imposed (kinematic) flow patterns it has been
suggested that there is no simple relation between $\alpha$ and helicity
of the flow; see \cite{CHT06}.
In order to clarify the existence of $\alpha$-effect and turbulent diffusion
as outlined above, and its $\Rm$ dependence, 
even in the kinematic limit, we have recently measured $\meanEMF$ directly
in numerical simulations of isotropic, homogeneous, 
helical turbulence \cite{sur_AB_KS07}. These simulations reach up to 
a modest $Rm\sim 220$. We find, somewhat surprisingly, that 
for isotropic homogeneous turbulence the high conductivity results 
obtained under FOSA are reasonably accurate up to the moderate 
values of $\Rm$ that we have tested.
A possible reason for this might be that the
predictions of FOSA are very similar to a closure 
called the minimal $\tau$ approximation (MTA) \cite{BF02,BS05a}
where the approximation of neglecting nonlinear terms made in FOSA
is not done, but replaced by a closure hypothesis.
But MTA is also not well justified, although numerical
simulations (for $\Rm\leq300$)
support some aspects of this closure \cite{BS05b}.
Interestingly, this agreement of $\alpha$ and $\eta_{\rm t}$ directly measured from
the simulation, with that expected under FOSA, is obtained even in the presence of a
small-scale dynamo, where $\bb$ is growing exponentially.
This suggests that the exponentially growing part of the
small-scale field does not make a contribution to the mean emf $\meanEMF$,
correlated with the mean field $\meanBB$.
It is essential to extend these results to even higher values of $\Rm$,
but these preliminary results are quite encouraging.

Another potential problem with the mean field dynamo paradigm 
is that magnetic helicity conservation puts severe restrictions 
on the strength of the $\alpha$-effect \cite{BS05a}.
The magnetic helicity associated with a field
${\bf B} = {\bf \nabla } \times {\bf A}$ is defined as
$H_T = \int {\bf A}.{\bf B} \ dV$, where ${\bf A}$ is the vector
potential \cite{Mof78}. Note that this definition of
helicity is only gauge invariant (and hence meaningful) if the domain
of integration is periodic, infinite or has
a boundary where the normal component of the field vanishes.
$H_T$ measures the linkages and twists in the magnetic field.
From the induction equation one can easily derive the
helicity conservation equation,
$dH_T/dt = -2 \eta \int \BB\cdot(\nab \times \BB) \ dV$. 
So in ideal MHD with $\eta = 0$, magnetic helicity
is strictly conserved, but this does not guarantee
conservation of $H_T$ in the limit $\eta\rightarrow0$.
For example magnetic energy, whose Ohmic dissipation is governed by
$(dE_B/dt)_{\rm Joule} = -\eta (4\pi/c^2) \int {\bf J}^2 dV$,
can be dissipated at finite rates
even in the limit $\eta\rightarrow0$, 
because small enough scales develop in the field (current sheets)
where the current density increases with decreasing $\eta$ as
$\propto\eta^{-1/2}$ as $\eta\rightarrow0$. 
Nevertheless, since helicity dissipation rate has a milder dependence on
the current density (its rate being proportional to $\JJ\cdot\BB$ and not
$\JJ^2$), in many astrophysical conditions where $R_m$ is
large ($\eta$ small), the magnetic helicity $H_T$, is almost independent
of time, even when the magnetic energy is dissipated at finite rates.

The operation of any mean-field dynamo automatically leads to 
the growth of linkages between the toroidal and poloidal mean fields
and hence a mean field helicity. In order to satisfy total
helicity conservation this implies that there must be equal and oppositely
signed helicity being generated in the fluctuating field.
What leads to this helicity transfer between scales?
To understand this, we need to split the helicity conservation
equation into evolution equations of the
sub-helicities associated with the mean field, say
$\overline{H}_T = \int \meanAA\cdot\meanBB \ dV$ and the fluctuating
field $h_T = \int \aaa\cdot\bb \ dV $.
The evolution equations for $\overline{H}_T$ and $h_T$ are \cite{BS05a}
\[
\frac{d\overline{H}_T}{dt} =\int 2 \meanEMF\cdot\meanBB \ dV
-2 \eta \int \frac{4\pi}{c} \meanJJ\cdot\meanBB \ dV
; \qquad \frac{dh_T}{dt} =-\int 2\meanEMF\cdot\meanBB \ dV
-2 \eta \int \frac{4\pi}{c} \jj\cdot\bb \ dV.
\]
Here, we have assumed that the surface terms can be taken to 
vanish (we will return to this issue below).
We see that the turbulent emf $\meanEMF$
transfers helicity between large and small scales;
it puts equal and opposite amounts of helicity into
the mean field and the small-scale field, conserving
the total helicity $H_T = \overline{H}_T + h_T$. 
Note that in the limit when the fluctuating field
has reached a stationary state one has $dh_T/dt \to 0$
and $\int \meanEMF\cdot\meanBB \ dV = 
-2 \eta \int (4\pi/ c) \jj\cdot\bb \ dV$ which tends to zero as $\eta \to 0$
for any reasonable spectrum of small scale current helicity.
Therefore, the component of the turbulent electromotive force
along the mean field is catastrophically quenched in the sense that
it tends to zero as $\eta \to 0$, or for large $\Rm$,
a feature also borne out in periodic box simulations \cite{B01,CH96}.

To make the above integral constraint into a local constraint
requires one to be able to define a gauge invariant helicity
{\it density} for at least the random small-scale field.
Such a definition has indeed been given, 
using the Gauss linking formula for helicity \cite{SB06}.
In physical terms, the magnetic helicity density $h$ 
of a random small scale field 
(in contrast to the total helicity $h_T$),
is the density of {\it correlated} links of the field.
This notion can be made precise and a {\it local} conservation
law can be derived (see \cite{SB06} for details) 
for the helicity density $h$,
\EQ
\frac{\partial h}{\partial t} + \nab\cdot\meanFF
= -2\meanEMF\cdot\meanBB-2\eta(4\pi/ c)\overline{\jj\cdot\bb}.
\label{finhel}
\EN
where $\meanFF$ gives a flux density of helicity.
(For a weakly inhomogeneous system, $h$ is approximately
$\overline{\aaa\cdot\bb}$ in the Coulomb gauge.)
In the presence of helicity fluxes,  we have
in the stationary limit, 
$\meanEMF\cdot\meanBB = -(1/2)\nab\cdot\meanFF
-\eta\overline{\jj\cdot\bb}$,
and therefore $\meanEMF\cdot\meanBB$ need not be 
catastrophically quenched.
Large scale dynamos then seem to need helicity fluxes
to work efficiently \cite{BF00,SSSB06,SSS07}.

This conclusion can be understood more physically as follows:
As the large-scale mean field grows the turbulent emf $\meanEMF$ is transferring
helicity between the small and large scale fields.
The large scale helicity is in the links of the mean poloidal
and toroidal fields, while the small scale helicity is
in what can be described as "twist" helicity of the 
small-scale field. Lorentz forces associated with the 
"twisted" small-scale field would like to untwist the field.
This would lead to an effective magnetic $\alpha$-effect
which opposes the kinetic $\alpha$ produced by the
helical turbulence (see below). The cancellation
of the total $\alpha$-effect is what leads to the
catastrophic quenching of the dynamo. This quenching can
be avoided if there is some way of transferring the twists
in the small scale field out of the region of dynamo action,
that is if there are helicity fluxes out of the system.

Blackman and Field (in \cite{BF00}) first suggested that the losses of the
small-scale magnetic helicity through the boundaries of the dynamo region can
be essential for mean-field dynamo action. Such a helicity flux can result
from the anisotropy of the turbulence combined with large-scale velocity shear
or the non-uniformity of the $\alpha$-effect \cite{BF00}.
Another type of helcity flux is simply advection
of the small scale field and its associated helicity out of the 
system, with $\meanFF = h \meanUU$ \cite{SSSB06}. 
This effect naturally arises in spiral galaxies where some of the gas
is heated by supernova explosions producing a hot phase that 
leaves the galactic disc, dragging along the small-scale part of
the interstellar magnetic field.

In order to examine the effect of helicity fluxes in more detail,
one also needs to fold in a model of how the dynamo
co-efficients get altered due to Lorentz forces.
Closure models either using the EDQNM closure 
or the MTA or quasi-linear theory,
suggest that the turbulent emf gets re-normalized, with
$\alpha = \alpha_K + \alpha_M$, where
$\alpha_M = (\tau/3) <{\bf b}.{\bf \nabla} \times {\bf b}>/(4\pi\rho)$,
is proportional to the small-scale current helicity
and is the magnetic alpha effect mentioned above \cite{BS05a,BF02,PFL76}.
The turbulent diffusion
$\eta_T$ is left unchanged to the lowest order, although to the next
order there arises a non-linear hyperdiffusive correction \cite{S99}.
Some authors have argued against an $\alpha_M$ contribution,
and suggest instead that the $\alpha$-effect
can be expressed exclusively in terms of
the velocity field, albeit one which is a solution
of the full momentum equation including the Lorentz force
\cite{proctor}. To clarify this issue, we have studied the nonlinear
$\alpha$-effect in the limit of small $\Rm,\Rey \ll 1$ using FOSA 
applied to both induction and momentum equations \cite{sur_KS_B07}.
We show explicitly in this limit that one can express $\alpha$ completely
in terms of the helical properties of the velocity field
as in traditional FOSA,
or, alternatively, as the sum of two terms,
a so-called kinetic $\alpha$-effect
and an oppositely signed term proportional to the
helical part of the small scale magnetic field,
akin to the above closures.

Adopting $\alpha = \alpha_K + \alpha_M$ as above, one 
can now look for a combined solution
to the helicity conservation equation (\ref{finhel}), and the mean-field
dynamo equation \cite{BS05a,BB02}, after relating the current helicity
arising in $\alpha_M$ to the magnetic helicity density $h$.
The effect of the advective flux in resolving the quenching of
the dynamo was worked out in detail in Ref \cite{SSSB06}.
In the absence of  an advective flux, the
initial growth of magnetic field is
catastrophically quenched and the large-scale magnetic field decreases at about
the same rate as it grew.
The initial growth occurs while the current helicity builds up to
cancel the kinetic $\alpha$-effect.
However, even a modest advective flux compensates the
catastrophic quenching of the dynamo and the mean field 
stays steady at about 10\%\ of the equipartition value.
(Excessive advection, however, hinders the dynamo as it
removes the mean field from the dynamo active region).
The vertical advection of magnetic helicity by galactic fountain flow
can therefore resolve in a straightforward fashion the
catastrophic quenching of nonlinear mean-field galactic dynamos.
Further, advection of small-scale magnetic fields 
may help the mean-field dynamo action in other ways.
For example, there is a concern that the back-reaction of Lorentz forces
due to the rapidly growing small-scale field could quench turbulent
transport processes, simply by suppressing the required Lagrangian chaos.
In this case their advection out of the galaxy
will still allow the dynamo to operate efficiently.

We have emphasized so far the role of helicity for the dynamo generation
of large-scale fields.
Intriguingly, several recent simulations show that the generation 
of large-scale fields may arise even 
in non-helical turbulence in the presence of strong enough 
shear \cite{B05}.
It is at present an important open question as the exact cause of such
large-scale field generation.

\section{Magnetizing the IGM through outflows}

We have already mentioned that outflows from supernovae and AGN could seed
galaxies and clusters with magnetic fields. We now consider whether such
processes could in fact magnetize the general intergalactic medium
and if so at what levels.

In this context it is interesting to note that there is 
growing evidence for metals in the moderately overdense
intergalactic medium (IGM) up to redshifts $z\sim 5$ or so 
(cf. \cite{AS07} and references therein).
The metals detected in the IGM can only have been synthesized by stars
in galaxies, and galactic outflows energized by supernovae (SNe) 
are the primary means by which they
can be transported from galaxies into the IGM. 
If galactic ISM is also magnetized due to some
form of dynamo action, then such outflows will also transport
magnetic fields into the IGM. One can
roughly estimate the IGM field resulting from such outflows, by
using magnetic flux conservation under spherically symmetric expansion;
that is $B_{\rm seed}\simeq(\rho_\mathrm{IGM}/\rho_\mathrm{ISM})^{2/3}
B_\mathrm{gal}$. For $ B_\mathrm{gal} \simeq 3 \mu$G, and
$\rho_\mathrm{IGM}/\rho_\mathrm{ISM} \simeq 10^{-6}$, one gets
$B_\mathrm{seed} \simeq 0.3$ nG.
In a similar fashion outflows from AGN
can also transport magnetized plasma into the IGM.

The level of such magnetization cannot be estimated
at present with any great certainty, since it depends 
not only on a host of parameters to do with the energetics of the outflows
but also on the evolution of the magnetic field in the outflow,
whether it is amplified by dynamo action or decays by generating
decaying MHD turbulence. Nevertheless there
are several attempts \cite{FL01}
where some model of the outflows is tied in with a model of
magnetic field behaviour. For example Bertone et al (in \cite{FL01})
find the outflows affected regions can have fields
ranging from $10^{-12} - 10^{-8}$ G, not much of a volume filling
at $z=3$ but a significant volume filling by $z=0$.
Due to outflows from AGN, Furlanetto and Loeb (in \cite{FL01}) estimate that 
by a redshift $z \sim 3$, about 5\%-20\% of the IGM volume is filled 
by magnetic fields and the magnetic pressure in these regions could
be comparable to the thermal pressure of the the photoionized IGM 
(at $T\sim 10^4$ K).

It should also be noted that such outflow scenarios are
constrained by the fact that they should not dynamically
perturb the IGM responsible for the Lyman-$\alpha$ forest
absorption at $z=3$. In this context
scenarios involving outflows from small mass galaxies
and filling the IGM at high redshifts are
more favored, for magnetizing the high $z$ IGM significantly
\cite{MFR01}, without perturbing the Lyman-$\alpha$ forest.

Of course, as mentioned earlier, one still needs to mix 
the magnetized and unmagnetized media and what this does to both 
the field strength and coherence scale is as yet unclear.
This would be relevant if we wish to use the IGM field to
explain galaxy and cluster magnetism.
Nevertheless,the magnetization of a significant volume of 
the IGM by $z=0$ could impact significantly on other
astrophysical phenomena. For example significant IGM 
magnetic fields would perturb cosmic ray propagation and 
is relevant to the issue of whether one can
do high-energy cosmic ray astronomy at all.

\section{Primordial magnetic fields from the early universe?}

We have so far concentrated on the hypothesis that
magnetic fields observed in galaxies and galaxy clusters
arise due to dynamo amplification of weak seed fields.
An interesting alternative is that the observed large-scale
magnetic fields are a relic from the
early Universe, arising perhaps during inflation or some other
phase transition (\cite{TW88} and references therein).
It is well known that scalar (density or potential)
perturbations and gravitational waves (or tensor perturbations)
can be generated during inflation. Could
magnetic field perturbations also be generated?

Indeed inflation provides several ideal conditions for the generation
of a primordial field with large coherence scales \cite{TW88}. 
First the rapid expansion in the inflationary era provides the kinematical
means to produce fields correlated on very large scales by just the
exponential stretching of wave modes. Also vacuum fluctuations
of the electromagnetic (or more correctly the hypermagnetic) field
can be excited while a mode is within 
the Hubble radius and these can be transformed
to classical fluctuations as it transits outside the Hubble radius.
Finally, during inflation any existing charged particle densities 
are diluted drastically by the expansion, so that the universe is 
not a good conductor; thus magnetic flux conservation then does not
exclude field generation from a zero field.
There is however one major difficulty; since the standard electromagnetic
action is conformally invariant, and the universe metric is conformally flat,
one can transform the evolution equation for the magnetic field
to its flat space version. The field then field always decreases
with expansion as $1/a^2$, where $a(t)$ is the expansion factor.

Therefore mechanisms for magnetic field generation need to invoke
the breaking of conformal invariance of the electromagnetic action,
which change the above behaviour to $B \sim 1/a^{\epsilon}$ with typically
$\epsilon \ll 1$ for getting a strong field. Since $a(t)$ is almost exponentially
increasing during slow roll inflation, the predicted field amplitude is
exponentially sensitive to any changes of the parameters of the model
which affects $\epsilon$. Therefore models of magnetic
field generation can lead to fields as large as $B\sim 10^{-9}$ G
(as redshifted to the present epoch) down to
fields which are much smaller than that required for even seeding the galactic dynamo.
Note that the amplitude of scalar perturbations generated
during inflation is also dependent on the parameters of the theory and
has to be fixed by hand. But the sensitivity to parameters seems
to be stronger for magnetic fields than for scalar perturbations due
to the above reason.

Another possibility is magnetic field generation in various phase transitions, like 
the electroweak transition or the QCD transition due to causal processes.
However these generically lead to a correlation scale of the field 
smaller than the Hubble radius at that epoch. Hence 
very tiny fields on galactic scales obtain, 
unless helicity is also generated; in which case
one can have an inverse cascade of energy to larger scales
\cite{BEO96}.

If a primordial magnetic field
with a present-day strength of even $B \sim 10^{-9}$~G
and coherent on Mpc scales is generated,
it can strongly influence a number of astrophysical processes. 
For example, such primordial magnetic fields could induce 
temperature and polarization anisotropies in
the Cosmic Microwave Background (CMB)
(see \cite{subramanian06} for a review). 
The signals that could be searched for include
excess temperature anisotropies (from scalar, vortical and tensor perturbations), 
B-mode polarization, and non-Gaussian
statistics \cite{Bcmb}. 
A field at a few nG level produces
temperature anisotropies at the $5\,\mu$K level, and B-mode polarization anisotropies
10 times smaller, and is therefore potentially detectable via the CMB anisotropies.
An even smaller field, with $B_0 \sim 0.1$ nG,
could lead to structure formation at high redshift $z > 15$, and hence
naturally impact on the re-ionization of the Universe \cite{SS05}.
A $0.1$nG field in the IGM could also
be sheared and amplified due to flux freezing,
during the collapse to form a galaxy
and lead to the few $\mu$G field observed in disk galaxies
(cf. \cite{kul}).
Ofcourse, one may still need a dynamo to maintain such a field
against decay and/or explain the observed global structure of
disk galaxy fields \cite{Shuk}.
Weaker primordial fields can also provide a strong
seed field for the dynamo. Overall, it is interesting to continue
to look for evidence of such a primordial field.

\section{Conclusions}

We have presented a brief overview of issues
related to the origin of cosmic magnetic fields. 
Battery mechanisms produce only a small seed field which needs 
to be amplified by a dynamo.
How fluctuation and mean-field dynamos 
work is at present under intense scrutiny. Basic ideas of turbulent dynamos
are in place. But a detailed understanding of the
saturation of fluctuation dynamos, and the linear/nonlinear
behaviour of turbulent transport co-efficients associated
with MFD's, is still challenging.
There is also increasing interest
in finding natural mechanisms for primordial field generation
in the early universe, and their observational consequences. 
Our knowledge of galactic,
cluster and IGM fields has come from observing the synchrotron polarization
and Faraday rotation in the radio wavelengths. Instruments like
the SKA will expand this knowledge enormously \cite{gaensler06}, and
will be crucial to further probe the magnetic universe.

\section{Acknowledgments}
I thank the SOC for providing support to attend this excellent meeting.


\begin{thebibliography}{99}

\bibitem{Beck00}
R.~Beck, \emph{Magnetic fields in normal galaxies}, Phil. Trans. of the Roy. Soc. of Lon., 
\textbf{A358}, 777--796 (2000);
T. E. Clarke, P. P. Kronberg, H. Bohringer,
\emph{A New Radio-X-Ray Probe of Galaxy Cluster Magnetic Fields}, ApJ Lett.,
\textbf{547}, L111--L114 (2001);
C. Vogt, T. A. En{\ss}lin, 
\emph{A Bayesian view on Faraday rotation maps - Seeing the magnetic
power spectra in galaxy clusters}, Astron. Astrophys., \textbf{434}, 67--76 (2005)

\bibitem{BS05a}
A.~{Brandenburg} and K.~{Subramanian}, \emph{{Astrophysical magnetic fields and
nonlinear dynamo theory}}, Phys. Rep.\, \textbf{417}, 1--209 (2005) [{\tt
arXiv:astro-ph/0405052}].

\bibitem{scheko05}
A. A. Schekochihin, S. C. Cowley, R. M. Kulsrud, G. W. Hammett and P. Sharma,
\emph{Plasma instabilities and magnetic field growth in clusters of galaxies}
Astrophys. J., \textbf{629}, 139--142 (2005)

\bibitem{Bier50}
L. Biermann, \emph{\"Uber den Ursprung der Magnetfelder auf
Sternen und im interstellaren Raum}, Z. Naturforsch.,
\textbf{5a}, 65--71 (1950); 
%
L. Mestel and I. W. Roxburgh, 
\emph{On the thermal generation of toroidal magnetic fields in
rotating stars}, Mon. Not. Roy. Astr. Soc., \textbf{136}, 615--626 (1962)


\bibitem{Sub94}
K. Subramanian, D. Narasimha and S. M. Chitre, \emph{Thermal generation 
of cosmological seed magnetic fields in ionization fronts}, 
Mon. Not. Roy. Astr. Soc. Lett., \textbf{271}, L15--L18 (2004)

\bibitem{gnedin00}
N. Y. Gnedin, A. Ferrara and E. G. Zweibel, 
\emph{Generation of the primordial magnetic fields during cosmological
reionization}, Astrophys. J., \textbf{539}, 505--516 (2000)

\bibitem{kulsrud97}
R. M. Kulsrud, R. Cen, J. P. Ostriker and D. Ryu, 
\emph{The protogalactic origin for cosmic magnetic fields}, 
Astrophys. J., \textbf{480}, 481--491 (1997);
G. Davies, L. M. Widrow,
\emph{A possible mechanism for generating galactic magnetic fields}, 
Astrophys. J., \textbf{540}, {755}{764} (2000)

\bibitem{harrison}
E. R. Harrison, 
\emph{Generation of magnetic fields in the radiation ERA},
Mon. Not. Roy. Astr. Soc., \textbf{147}, 279--286 (1970);
I. N. Mishustin and A. A. Ruzmaikin,
\emph{Occurrence of 'priming' magnetic fields during the formation
of protogalaxies}, JETP, \textbf{61}, 441--444 (1971), [Sov. Phys. JETP,
\textbf{34}, 233--235 (1972)]

\bibitem{gopal-sethi}
R. Gopal and S. Sethi,
\emph{Generation of magnetic field in the pre-recombination era},
Mon. Not. Roy. Astr. Soc., \textbf{363} 521-528 (2005);
S. Matarrese, S. Mollerach, A. Notari and A. Riotto,
\emph{Large-scale magnetic fields from density perturbations}
Phys. Rev. D. \textbf{71}, 043502-1--043502-7 (2005);
Kiyotomo Ichiki, Keitaro Takahashi, Naoshi Sugiyama, Hidekazu Hanayama, Hiroshi Ohno,
\emph{Magnetic field spectrum at cosmological recombination},
[{\tt   arXiv:astro-ph/0701329}]

\bibitem{Rees87}
M. J. Rees, \emph{The origin and cosmogonic implications of seed magnetic fields}, 
Quart. J. Roy. Astr. Soc., \textbf{28}, 197--206 (1987);
M. J. Rees, \emph{Origin of cosmic magnetic fields}, 
Astron.\ Nachr.\, \textbf{327}, 395--398 (2006)

\bibitem{RSS88}
A. A. Ruzmaikin, D. D. Sokoloff and A. M. Shukurov,
\emph{Magnetic Fields of Galaxies}, Kluwer, Dordrecht (1988)

\bibitem{Kaz68}
A. P. Kazantsev, \emph{Enhancement of a magnetic field by a 
conducting fluid}, JETP, \textbf{26}, 1031--1034 (1968)

\bibitem{Zel90}
Y. B. Zeldovich, A. A. Ruzmaikin and D. D. Sokoloff, 
\emph{The Almighty Chance}, World Scientific, Singapore (1990)

\bibitem{Haugen}
N. E. L. Haugen, A. Brandenburg, W. Dobler, 
\emph{Is nonhelical hydromagnetic turbulence peaked at small scales?},
Astrophys. J. Lett., \textbf{597}, L141--L144 (2003);
N. E. L. Haugen, A. Brandenburg, W. Dobler, 
\emph{Simulations of nonhelical hydromagnetic turbulence},
Phys. Rev. E., \textbf{70}, 016308-1--016308-14 (2004)

\bibitem{Schek04} 
A. A. Schekochihin, S. C. Cowley, S. F. Taylor,
J. L. Maron, J. C. McWilliams, 
\emph{Simulations of the small scale turbulent dynamo},
Astrophys. J., \textbf{612}, 276--307 (2004)

\bibitem{S99}
K. Subramanian, \emph{Unified treatment of small and large scale
dynamos in helical turbulence}, Phys. Rev. Lett., \textbf{83}, 2957--2960 (1999);
K. Subramanian, 
\emph{Hyperdiffusion in nonlinear, large and small scale turbulent dynamos}
Phys. Rev. Lett., \textbf{90}, 245003-1--245003-4 (2003)

\bibitem{SSH06}
K. Subramanian, A. Shukurov and N. E. L  Haugen,
\emph{Evolving turbulence and magnetic fields in galaxy clusters},
Mon. Not. Roy. Astr. Soc., \textbf{366}, 1437-1454 (2006)

\bibitem{vogt_ensslin06}
C. Vogt, T. A. En{\ss}lin, 
\emph{Magnetic turbulence in cool cores of galaxy clusters},
Astron. Astrophys., \textbf{453}, 447--458 (2006)

\bibitem{B01} 
A. Brandenburg, \emph{The inverse cascade and nonlinear alpha effect in simulations
of isotropic helical hydromagnetic turbulence}, 
Astrophys. J., \textbf{550}, 824--840 (2001)

\bibitem{Par55}
E. N. Parker, 
\emph{Hydromagnetic dynamo models}, 
Astrophys. J., \textbf{122}, 293--314 (1955);

\bibitem{Mof78}
H. K. Moffatt,
\emph{Magnetic field generation in electrically conducting fluids}, 
Cambridge University Press, Cambridge (1978);
F. Krause, K.-H. R\"adler,
\emph{Mean-Field Magnetohydrodynamics and Dynamo Theory}
Akademie-Verlag, Berlin; also Pergamon Press, Oxford (1980)

\bibitem{Shuk}
A. Shukurov,
\emph{Introduction to galactic dynamos},
in "Mathematical Aspects of Natural Dynamos",eds. E. Dormy and
B. Desjardins, EDP Press (2004)
[{\tt arXiv:astro-ph/0411739}]

\bibitem{CHT06}
A. Courvoisier A., D. W. Hughes and S. M. Tobias,
\emph{$\alpha$ effect in a family of chaotic flows},
Phys. Rev. Lett., \textbf{96}, 034503-1--034503-4 (2006)

\bibitem{sur_AB_KS07}
S. Sur, A. Brandenburg and K. Subramanian,
\emph{Kinematic alpha effect in isotropic turbulence simulations},
Mon. Not. Roy. Astr. Soc. (in press), (2007) [{\tt arXiv:0711.3789}]

\bibitem{CH96}
F. Cattaneo and D. W. Hughes,
\emph{Nonlinear saturation of the turbulent $\alpha$ effect},
Phys. Rev. E., \textbf{54}, R4532--R4535 (1996)

\bibitem{SB06}
K. Subramanian and A. Brandenburg, \emph{Magnetic helicity density and its 
flux in weakly inhomogeneous turbulence}, Astrophys. J. Lett., \textbf{648},
L71--L74 (2006) [{\tt astro-ph/0509392~v1 contains more details}]

\bibitem{BF02}
E. G. Blackman, G. B. Field, 
\emph{New dynamical mean field dynamo theory and closure approach}, 
Phys. Rev. Lett., \textbf{89}, 265007-1--265007-4 (2002);
K.-H. R\"adler, N. Kleeorin, I. Rogachevskii,
\emph{The mean electromotive force for MHD turbulence: the case of
a weak mean magnetic field and slow rotation}, Geophys. Astropys. Fluid. Dyn.,
\textbf{97} 249--274 (2003)

\bibitem{BS05b}
A.~{Brandenburg} and K.~{Subramanian}, 
\emph{Minimal tau approximation and simulations of the alpha effect},
Astron. Astrophys., \textbf{439}, 835--843 (2005);
A.~{Brandenburg} and K.~{Subramanian},
\emph{Simulations of the anisotropic kinetic and magnetic alpha effects},
Astron.\ Nachr.\, \textbf{328}, 507--512 (2007)


\bibitem{BF00}
E. G. Blackman and G. B. Field, 
\emph{Constraints on the magnitude of $\alpha$ in dynamo Theory},
Astrophys. J., \textbf{534}, 984--988 (2000);
N. Kleeorin, D. Moss, I. Rogachevskii and D. Sokoloff,
\emph{Helicity balance and steady-state strength of 
the dynamo generated galactic magnetic field},
Astron. Astrophys. Lett., \textbf{361}, L5-L8 (2000);
E. T. Vishniac and J. Cho, 
\emph{Magnetic Helicity Conservation and Astrophysical Dynamos},
Astrophys. J., \textbf{550}, 752--760 (2001);
A.~{Brandenburg} and K.~{Subramanian},
\emph{Strong mean field dynamos require supercritical helicity fluxes},
Astron.\ Nachr.\, \textbf{326},400--408 (2005);
K. Subramanian and A. Brandenburg, 
\emph{Nonlinear current helicity fluxes in turbulent dynamos and alpha quenching},
Phys. Rev. Lett., \textbf{93}, 205001-1--205001-4 (2004)

\bibitem{SSSB06}
A. Shukurov, D. Sokoloff, K. Subramanian and A. Brandenburg,
\emph{Galactic dynamo and helicity losses through fountain flow},
Astron. Astrophys. Lett., \textbf{448}, L33--L36 (2006)

\bibitem{SSS07}
S. Sur, A. Shukurov and K. Subramanian, 
\emph{Galactic dynamos supported by magnetic helicity fluxes},
Mon. Not. Roy. Astr. Soc., \textbf{377}, 874--882 (2007)

\bibitem{PFL76}
A.~Pouquet, U.~Frisch, J.~L\'eorat,
\emph{Strong MHD helical turbulence and the nonlinear dynamo effect},
Jour. of Fluid Mech., \textbf{77}, 321-354 (1976);
A. V. Gruzinov, P. H. Diamond,
\emph{Self-consistent theory of mean field electrodynamics},
Phys. Rev. Lett., \textbf{72}, 1651--1653 (1994)

\bibitem{proctor}
M. R. E. Proctor,
\emph{Dynamo processes: the interaction of turbulence and magnetic fields},
In Stellar Astrophysical Fluid Dynamics, eds. M. J. Thompson, J. Christensen-Dalsgaard,
Cambridge University Press, 143--158 (2003);\  
K.-H. R\"adler and M. Rheinhardt,
\emph{Mean-field electrodynamics: critical analysis of 
various analytical approaches to the mean electromotive force}
Geophys. Astropys. Fluid. mech., \textbf{101}, 117--154 (2007)

\bibitem{sur_KS_B07}
S. Sur, K. Subramanian and A. Brandenburg,
\emph{Kinetic and magnetic $\alpha$-effects in non-linear dynamo theory},
Mon. Not. Roy. Astr. Soc., \textbf{376}, 1238--1250 (2007)

\bibitem{BB02}
E. G. Blackman and A. Brandenburg,
\emph{Dynamic nonlinearity in large scale dynamos with shear},
Astrophys. J., \textbf{579}, 359--373 (2002);
K. Subramanian, 
\emph{Magnetic helicity in galactic dynamos}, 
Bull.\ Astr.\ Soc.\ India, \textbf{30}, 715--721 (2002)

\bibitem{B05}
A. Brandenburg,
\emph{The case for a distributed solar dynamo shaped by near-surface shear},
Astrophys. J., \textbf{625}, 539--547 (2005);
A. Brandenburg, K.-H. R\"adler, M. Rheinhardt and
P. J. K\"apyl\"a, \emph{Magnetic diffusivity tensor and dynamo 
effects in rotating and shearing turbulence},
Astrophys. J. (in press), [{\tt arXiv:0710.4059}];
T. A. Yousef et al.,
\emph{Generation of magnetic field by combined action of turbulence and shear}
[{\tt 	arXiv:0710.3359}]

\bibitem{AS07}
A. Aguirre and J. Schaye,
\emph{How did the IGM become enriched?},
EAS Publications Series, \textbf{24}, 165--175 (2007) [{\tt arXiv:astro-ph/0611637v1}]

\bibitem{FL01}
S. R. Furlanetto and A. Loeb, 
\emph{Intergalactic magnetic fields from quasar outflows}
Astrophys. J., \textbf{556}, 619--634 (2001);
P. P. Kronberg, H. Lesch and U. Hopp,
\emph{Magnetization of the intergalactic medium by primeval galaxies},
Astrophys. J., \textbf{511}, 56--64 (1999);
S. Bertone, C. Vogt and T. En{\ss}lin,
\emph{Magnetic field seeding by galactic winds},
Mon. Not. Roy. Astr. Soc., \textbf{370}, 319--330 (2006)

\bibitem{MFR01}
P. Madau, A. Ferrara and M. J. Rees, 
\emph{Early Metal Enrichment of the Intergalactic Medium by Pregalactic Outflows},
Astrophy. J., \textbf{555}, 92--105 (2001);
S. Samui, K. Subramanian and R. Srianand,
\emph{Constrained semi-analytical models of Galactic outflows},
Mon. Not. Roy. Astr. Soc. (in Press) (2008)[{\tt arXiv:0801.1401}]

\bibitem{TW88} 
M. Turner, L. M. Widrow,
\emph{Inflation-produced, large scale magnetic fields}, 
Phys. Rev. D. \textbf{37}, 2743--2754 (1988);
L. M. Widrow, \emph{Origin of galactic and extragalactic magnetic fields}, 
Rev. Mod. Phys., \textbf{74}, 775--823 (2002);
M. Giovannini,
\emph{Magnetic fields, strings and cosmology},
To appear in the book "String theory and fundamental interactions",
eds. M. Gasperini and J. Maharana, Lecture Notes in Physics, Springer,
Berlin/Heidelberg (2007) [{\tt 	arXiv:astro-ph/0612378}]

\bibitem{BEO96}
A. Brandenburg, K. Enqvist, P. Olesen,
\emph{Large-scale magnetic fields from hydromagnetic turbulence
in the very early universe}, Phys. Rev. D., \textbf{54}, 1291--1300 (1996);
R. Banerjee and K. Jedamzik, 
\emph{Evolution of cosmic magnetic fields: 
From the very early Universe, to recombination, to the present},
Phys. Rev. D., \textbf{70}, 123003-1--123003-25 (2004)

\bibitem{subramanian06}
K. Subramanian,
\emph{Primordial magnetic fields and CMB anisotropies},
Astron. Nachr., \textbf{327}, 403--409 (2006);

\bibitem{Bcmb}
J. D. Barrow, P. G. Ferreira and J. Silk,
\emph{Constraints on a Primordial Magnetic Field},
Phys. Rev. Lett., \textbf{78}, 3610--3613 (1997); \
K. Subramanian and J. D. Barrow, 
\emph{Microwave Background Signals from Tangled Magnetic Fields}, 
Phys. Rev. Lett., \textbf{81}, 3575--3578 (1998); \
T. R. Seshadri and K. Subramanian, 
\emph{Cosmic Microwave Background Polarization Signals from Tangled
Magnetic Fields},
Phys. Rev. Lett., \textbf{87}, 101301-1--101301-4 (2001); \
A. Mack, T. Kahniashvili and A. Kosowsky, 
\emph{Microwave background signatures of a primordial stochastic magnetic
field},
Phys. Rev. D., \textbf{65}, 123004 (2002); \
K. Subramanian, T. R. Seshadri and J. D. Barrow, 
\emph{Small-scale cosmic microwave background polarization anisotropies
due to tangled primordial magnetic fields}
Mon. Not. Roy. Astr. Soc., \textbf{344}, L31--L35 (2003); \
A. Lewis, 
\emph{CMB anisotropies from primordial inhomogeneous magnetic fields}, 
Phys. Rev. D., \textbf{70}, 043011 (2004); \
M. Giovannini,
\emph{Tight coupling expansion and fully inhomogeneous magnetic fields},
Phys. Rev. D., \textbf{74}, 063002 (2006)

\bibitem{SS05}
S. Sethi and K. Subramanian,
\emph{Primordial magnetic fields in the post-recombination era and early reionization},
Mon. Not. Roy. Astr. Soc., \textbf{356}, 778--788 (2005); \
H. Tashiro and N. Sugiyama,
\emph{Early reionization with primordial magnetic fields},
Mon. Not. Roy. Astr. Soc., \textbf{368}, 965--970 (2006)

\bibitem{kul}
R. M. Kulsrud and E. G. Zweibel,
\emph{The origin of astrophysical magnetic fields},
[{\tt arXiv:0707.2783}]

\bibitem{gaensler06}
B. M. Gaensler,
\emph{The Square Kilometre Array: a new probe of cosmic magnetism},
Astron.\ Nachr.\, \textbf{327}, 387--394 (2006); \
B. M. Gaensler, R. Beck and L. Feretti,
\emph{The origin and evolution of cosmic magnetism},
New Astron. Rev., \textbf{48}, 1003--1012 (2004)

\end{thebibliography}
\end{document}